\pdfoutput=1

\documentclass[11pt,twoside,a4paper,cmspaper,final,collab]{cms-tdr}
\def\svnVersion{450509P}\def\cmsCernNoTag{CERN-EP-2017-232}\def\cmsCernDate{\today}\def\cmsMessage{Published in Physical Review Letters as \href{http://dx.doi.org/10.1103/PhysRevLett.120.081801}{\doi{10.1103/PhysRevLett.120.081801.}}}

\begin{document}\cmsNoteHeader{SMP-17-004}

\hyphenation{had-ron-i-za-tion}
\hyphenation{cal-or-i-me-ter}
\hyphenation{de-vices}
\RCS$Revision: 440801 $
\RCS$HeadURL: svn+ssh://svn.cern.ch/reps/tdr2/papers/SMP-17-004/trunk/SMP-17-004.tex $
\RCS$Id: SMP-17-004.tex 440801 2018-01-09 21:18:17Z alverson $

\newcommand{\mll}{\ensuremath{m_{\ell\ell}}}
\newcommand{\dyll}{\ensuremath{\Z/\gamma^*\to\ell^+\ell^-}}

\newlength\cmsFigWidth
\ifthenelse{\boolean{cms@external}}{\setlength\cmsFigWidth{0.85\columnwidth}}{\setlength\cmsFigWidth{0.4\textwidth}}
\ifthenelse{\boolean{cms@external}}{\providecommand{\cmsLeft}{top\xspace}}{\providecommand{\cmsLeft}{left\xspace}}
\ifthenelse{\boolean{cms@external}}{\providecommand{\cmsRight}{bottom\xspace}}{\providecommand{\cmsRight}{right\xspace}}
\ifthenelse{\boolean{cms@external}}{\providecommand{\NA}{\ensuremath{\cdots}\xspace}}{\providecommand{\NA}{\ensuremath{\text{---}}\xspace}}
\ifthenelse{\boolean{cms@external}}{\providecommand{\cmsFigOneABC}{upper-left, upper-right, lower-left}}{\providecommand{\cmsFigOneABC}{left, middle-left, middle-right}}
\ifthenelse{\boolean{cms@external}}{\providecommand{\cmsFigOneD}{lower-right}}{\providecommand{\cmsFigOneD}{right}}
\newcommand{\x}{\ensuremath{\phantom{0}}}
\newcommand{\y}{\ensuremath{\phantom{.}}}
\ifthenelse{\boolean{cms@external}}
{\providecommand{\suppMaterial}{the supplemental material~\cite{supp}, which includes Ref.~\cite{Zaro:2002500},}}
{\providecommand{\suppMaterial}{Appendix~\ref{app:suppMat}}}

\cmsNoteHeader{SMP-17-004}
\title{Observation of electroweak production of same-sign \texorpdfstring{$\PW$}{W} boson pairs in the two jet and two same-sign lepton final state in proton-proton collisions at \texorpdfstring{13\TeV}{13 TeV}}

\date{\today}

\abstract{
The first observation of electroweak production of same-sign $\PW$ boson pairs in proton-proton collisions is reported. The data sample corresponds to an integrated luminosity of 35.9\fbinv collected at a center-of-mass energy of 13\TeV with the CMS detector at the LHC. Events are selected by requiring exactly two leptons (electrons or muons) of the same charge, moderate missing transverse momentum, and two jets with a large rapidity separation and a large dijet mass. The observed significance of the signal is 5.5 standard deviations, where a significance of 5.7 standard deviations is expected based on the standard model. The ratio of measured event yields to that expected from the standard model at leading-order is $0.90 \pm 0.22$. A cross section measurement in a fiducial region is reported.  Bounds are given on the structure of quartic vector boson interactions in the framework of dimension-eight effective field theory operators and on the production of doubly charged Higgs bosons.
}

\hypersetup{%
pdfauthor={CMS Collaboration},%
pdftitle={Observation of electroweak production of same-sign W boson pairs in the two jet and two same-sign lepton final state in proton-proton collisions at 13 TeV},%
pdfsubject={CMS},%
pdfkeywords={CMS, physics, vector boson scattering}}

\maketitle
The standard model (SM) of particle physics provides an accurate description
of observations from many accelerator- and nonaccelerator-based
experiments. The discovery of a Higgs
boson~\cite{Aad:2012tfa,Chatrchyan:2012xdj,CMSPaperCombinationLong} confirmed
that $\PW$ and $\PZ$ gauge bosons acquire mass using the Higgs
mechanism. This discovery motivates further study of
the mechanism of electroweak (EW) symmetry breaking through measurements of
vector boson scattering (VBS)  processes. Physics models beyond the SM predict enhancements
in VBS through modifications of the Higgs sector or the presence of
additional resonances~\cite{PhysRevD.87.055017, PhysRevD.87.093005}.

The main goal of this analysis is to identify same-sign $\PW$ boson
pairs produced in association with two jets purely via the electroweak interaction.
Candidate events contain exactly two identified leptons (electrons or muons) of the
same charge, moderate missing transverse momentum ($\ptmiss$), and two jets
with a large rapidity separation and a large dijet mass. The selection
of same-sign lepton events reduces the contribution from the strong
production of $\PW$ boson pairs, making the
experimental signature an ideal topology for VBS studies.

\begin{figure}[htb!]
\centering
\includegraphics[width=0.23\textwidth]{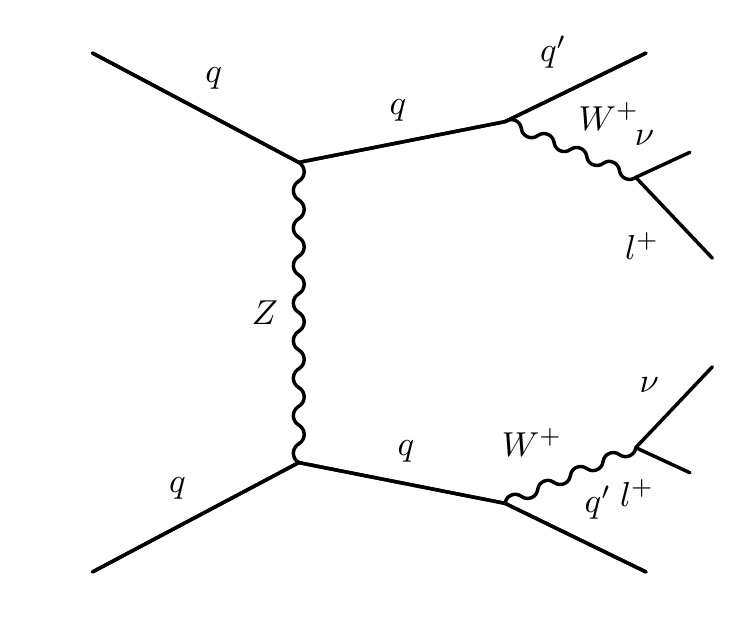}
\includegraphics[width=0.23\textwidth]{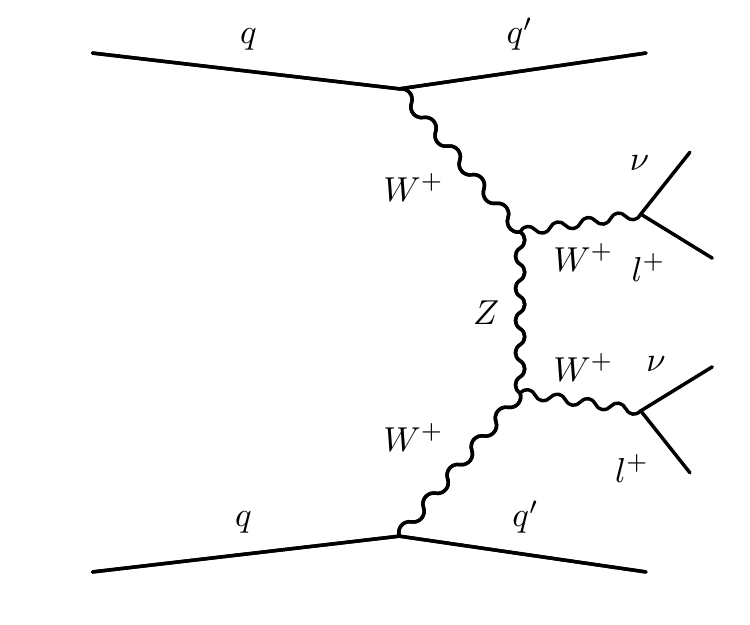}
\includegraphics[width=0.23\textwidth]{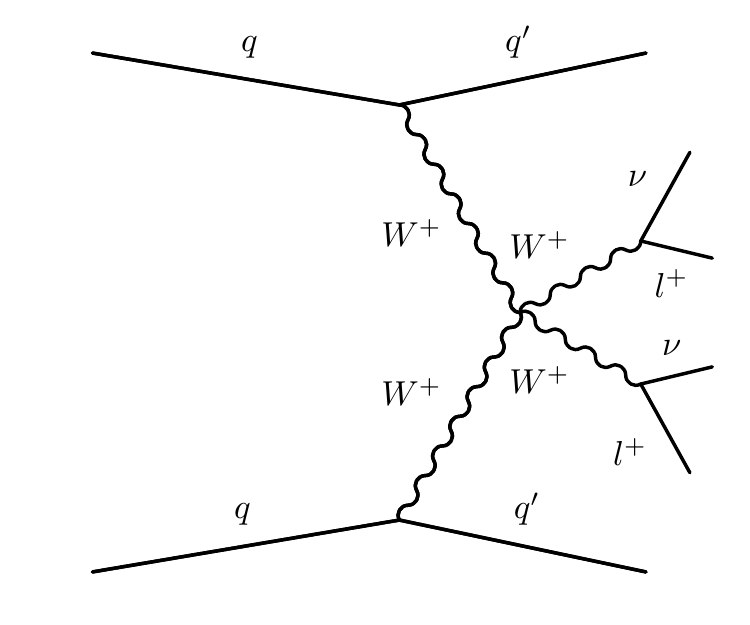}
\includegraphics[width=0.23\textwidth]{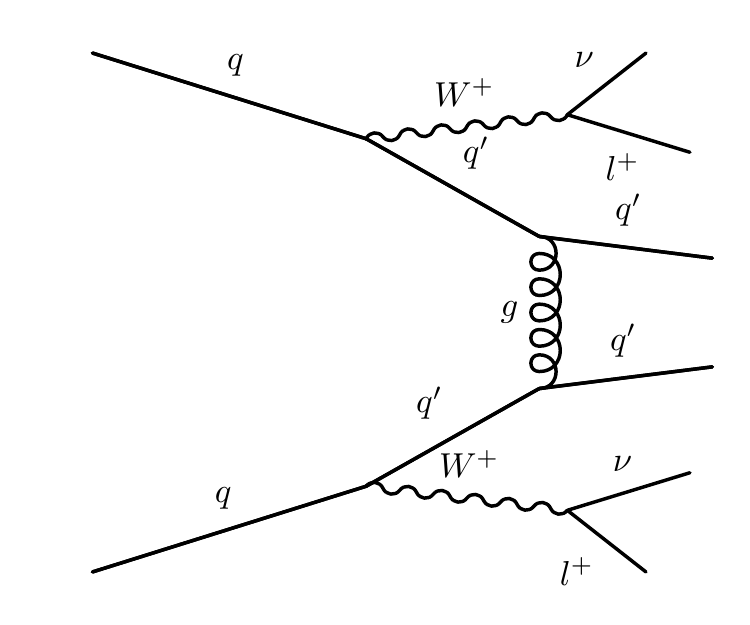}
\caption{Representative Feynman diagrams for single, triple, and
quartic gauge couplings of the EW-induced same-sign $\PW$ boson pair
production (\cmsFigOneABC) and QCD-induced background (\cmsFigOneD).}
\label{fig:diagrams}
\end{figure}

Figure~\ref{fig:diagrams} shows representative Feynman diagrams for EW and quantum chromodynamics (QCD) induced same-sign $\PW$ boson pair production.

An excess of events with respect to SM expectation
could signal the presence of anomalous quartic gauge
couplings (aQGC)~\cite{aqgc_operators} or the existence of a new
resonance, such as a doubly charged Higgs boson. Doubly charged Higgs bosons are predicted in Higgs sectors beyond the SM where weak isotriplet
scalars are included~\cite{CE1, CE2}. They can be produced via
vector-boson fusion (VBF) and decay to pairs of same-sign $\PW$ bosons~\cite{1202.2014}.

First experimental results for EW same-sign $\PW$ bosons searches were
reported by the ATLAS and CMS Collaborations
based on data collected at $\sqrt{s}=8\TeV$, corresponding to an integrated luminosity of approximately
20\fbinv~\cite{PhysRevLett.113.141803, Khachatryan:2014sta}.
The observed significance is 3.6 (2.0) standard deviations
for the ATLAS (CMS) study, where a significance of 2.8 (3.1) standard
deviations is expected based on the SM prediction.
This Letter presents a study of VBS in the same-sign $\PW$ boson pair
final state at $\sqrt{s}=13\TeV$. The data sample corresponds to an
integrated luminosity of $35.9 \pm 0.9\fbinv$ collected with the CMS
detector~\cite{Chatrchyan:2008zzk} at the CERN LHC in 2016.

The central feature of the CMS apparatus is a superconducting solenoid
of 6\unit{m} internal diameter, providing a magnetic field of
3.8\unit{T}. Within the solenoid volume are a silicon pixel and strip
tracker, a lead tungstate crystal electromagnetic calorimeter,
and a brass and scintillator hadron calorimeter, each composed
of a barrel and two endcap sections. Forward calorimeters extend the
pseudorapidity $\eta$ coverage provided by the barrel and endcap
detectors. Muons are detected in gas-ionization detectors embedded in
the steel flux-return yoke outside the solenoid. A more detailed
description of the CMS detector, together with a definition of the
coordinate system used and the relevant kinematic variables, can be
found in Ref.~\cite{Chatrchyan:2008zzk}.

The signal and background processes are simulated using the Monte
Carlo (MC) generator \MADGRAPH{}5\_a\MCATNLO~2.3.3~\cite{Alwall:2014}.
The same-sign $\PW$ boson pair samples are produced at leading-order (LO) via diagrams with two or fewer
QCD and up to six EW vertices. This includes
two categories of diagrams: those with exactly two QCD vertices, which
we refer to as QCD production, and those with no QCD vertices, which we refer to
as EW production. We consider only the EW production as the
signal in the analysis, whereas the QCD production is considered as
background.  This background is small and can be kinematically
separated from the signal.
The interference between the EW and QCD processes
is at the level of a few percent in the signal region and is treated as a systematic
uncertainty. The EW category includes diagrams with $\PW\PW\PW\PW$
quartic interactions and diagrams where two same-sign $\PW$ bosons
scatter through the exchange of a Higgs boson, a $\PZ$ boson,
or a photon. The $\PW\PZ$ and $\PZ\PZ$ production processes via
$\PQq\PAQq$ annihilation and the $\PW\PGg$ process are generated at LO.
The Drell--Yan, $\PZ\PGg$, $\ttbar$, $\ttbar \PW$, $\ttbar \PZ$, $\PW\PZ\PZ$,
$\PW\PW\PZ$, $\PW\PW\PW$, and $\PZ\PZ\PZ$ samples are generated at
next-to-leading order (NLO). The simulated samples of background
processes are normalized to the best theoretical prediction. The \PYTHIA~8.205~\cite{Sjostrand:2015} package is
used  for parton showering, hadronization, and the underlying event
simulation, with tune CUETP8M1~\cite{Skands:2014pea,Khachatryan:2015pea}.
The NNPDF~3.0~\cite{nnpdf} set is used as the default set of parton
distribution functions.
The detector response is simulated by the \GEANTfour
package~\cite{Agostinelli:2002hh}, based on a detailed description of
the CMS detector.
Proton-proton interactions occurring in the same beam crossing bin as
the event of interest (pileup) are included in the simulation samples.
The simulated pileup has a mean of approximately 27 and corresponds to the conditions observed in the 13\TeV data collected in 2016.

The final states considered are
$\Pep \Pep \PGn_{\Pe} \PGn_{\Pe} j j$,
$\Pep \Pgmp \PGn_{\Pe} \PGn_{\Pgm} j j$,
$\Pgmp\Pgmp \PGn_{\Pgm} \PGn_{\Pgm} j j$,
and their charge conjugates. The electrons
and muons can be directly produced from a $\PW$ boson decay or from a
$\PW$ boson with an intermediate $\PGt$ lepton decay.
A suite of single- and double-lepton triggers
are used for this analysis~\cite{Khachatryan:2016bia}. The trigger efficiency for the signal
process is larger than 99.8\% after all other selection
requirements are applied.

{\tolerance=700
A particle-flow algorithm~\cite{Sirunyan:2017ulk} is used to reconstruct
observable particles in the event. It combines all subdetector information
to reconstruct individual particles and identify them as charged and neutral hadrons, photons, and leptons.
Electrons and muons are reconstructed by associating a track reconstructed in
the silicon detectors either with a cluster of energy in the electromagnetic
calorimeter~\cite{Khachatryan:2015hwa} or a track in the muon system.
The $\ptmiss$ is defined as the magnitude of the negative vector sum of the transverse momenta of all reconstructed
particles (charged and neutral) in the event modified by corrections
to the energy scale of reconstructed jets.
\par}

The event selection aims to identify same-sign lepton events with
the VBS topology, while reducing the contribution from QCD-induced same-sign $\PW$ boson pairs, the top quark,
Drell--Yan, and $\PW\PZ$ background contributions.
Two same-sign leptons, electrons or muons, with transverse
momentum $\pt > 25\,(20)$\GeV for the leading (trailing) lepton and $\abs{\eta}<2.5\,(2.4)$ for  electrons (muons) are required.
Electrons and muons are required to be isolated from other charged and neutral particles in the event.
Jets are reconstructed using the anti-\kt
clustering algorithm~\cite{antikt}  with a distance parameter $R=0.4$,
as implemented in the \textsc{fastjet}  package~\cite{Cacciari:fastjet1,Cacciari:fastjet2}.
Events are required to contain at least two jets with
$\pt>30\GeV$ and $\abs{\eta}< 5.0$.
The VBS topology is targeted by requiring that the
two highest \pt jets have a large dijet mass, $m_{jj} > 500\GeV$, a large $\eta$
separation, $\abs{\Delta \eta_{jj}} > 2.5$, and $\max(z_{\ell}^{*}) < 0.75$,
where $z_{\ell}^{*} = \abs{\eta_{\ell}-(\eta_{j1}+\eta_{j2})/2}/\abs{\Delta \eta_{jj}}$ is
the Zeppenfeld variable~\cite{Rainwater:1996ud}, $\eta_{\ell}$ is the
pseudorapidity of a lepton, and $\eta_{j1}$ and $\eta_{j2}$ are the
pseudorapidities of the leading and subleading jet, respectively.

Techniques for identification of b quark jets are used to veto
top quark events.
These techniques are based on b quark jet tagging criteria that
combine the information from displaced tracks with the information from secondary vertices
associated with the jet using a multivariate technique, and on the
possible presence of a
soft muon in the event from the semileptonic decay of the bottom
quark~\cite{btag,BTV-15-001}. A minimum
dilepton mass,  $\mll > 20\GeV$, is required to reduce nonprompt
lepton processes.  To reduce the background from $\PW\PZ$ production, events with a third loosely identified
lepton with $\pt > 10\GeV$ or an identified hadronically-decaying $\tau$ lepton with $\pt > 18\GeV$ are rejected.
Drell--Yan events can be selected if the charge of one lepton is
measured incorrectly. The charge mismeasurement in dimuon events is
negligible, while this background is not negligible for dielectron events.
An invariant mass veto, $\abs{\mll - m_\PZ} > 15\GeV$, is imposed for
$\Pe^{\pm} \Pe^{\pm}$ events. The  Drell--Yan background is further reduced by requiring $\ptmiss > 40\GeV$.

A $\PW\PZ\to 3\ell\PGn$ control region is defined by requiring an
additional identified lepton with $\pt > 10\GeV$ and an opposite-sign same-flavor lepton pair with an
invariant mass consistent with that of the $\PZ$ boson.
The background contribution from charge misidentification is estimated
by applying a data-to-simulation efficiency correction to charge misidentified
electrons in bins of $\eta$.
The charge misidentification rate, estimated using Drell--Yan events, is between about 0.01\% in the barrel region and about 0.3\% in the endcap
region for electrons.

The nonprompt lepton backgrounds originating from leptonic decays of
heavy quarks, hadrons mis\-identi\-fi\-ed as leptons, and electrons
from photon conversions are suppressed by the identification and
isolation requirements imposed on electrons and muons.
The remaining contribution from the nonprompt lepton background is
estimated directly from data following the technique described in
Ref.~\cite{Khachatryan:2014sta}. All other background processes are
estimated from simulation applying corrections to account for small
differences between data and simulation, as described below.

The lepton trigger, reconstruction, and selection efficiencies are
measured using Drell--Yan events that provide an
unbiased sample with high purity. The estimated uncertainty
is less than 2\% per lepton.
The jet energy
scale and resolution uncertainties give rise to an uncertainty in the
yields of up to 7\%.
The uncertainty in the estimated event yields related to the
top quark veto is evaluated
by using a $\dyll$ sample with at least two reconstructed jets
and is 3\% or smaller. The statistical uncertainty due to the finite size of each simulated sample is also taken into account. The uncertainty of 2.5\% in the integrated
luminosity determination~\cite{LUM-17-001} is considered for all processes estimated
from simulation and for the fiducial cross section.
The normalization of the processes with misidentified leptons is
estimated with a systematic uncertainty of 30\%.
The $\PW\PZ$ background normalization
uncertainty is 20--40\%, dominated by the statistical uncertainty arising from the small number of events in the
trilepton control region. Theoretical uncertainties are estimated by varying simultaneously
the renormalization and factorization scales up and down by a factor of two from
their nominal value in each event, and, depending
on kinematic region, are up to 12\% for the signal
normalization and 20\% for the triboson background normalization.
The interference between the EW signal and the QCD-induced same-sign
$\PW$ boson production background is estimated using the \textsc{Phantom}~1.2.8
generator~\cite{Ballestrero:2007xq} and is treated as a systematic
uncertainty of 4.5\% in the signal yield.
An uncertainty in the parton distribution function contributes 5\% to
the signal times acceptance~\cite{Butterworth:2015oua}.

\begin{figure*}[htb]
\centering
\includegraphics[width=\cmsFigWidth]{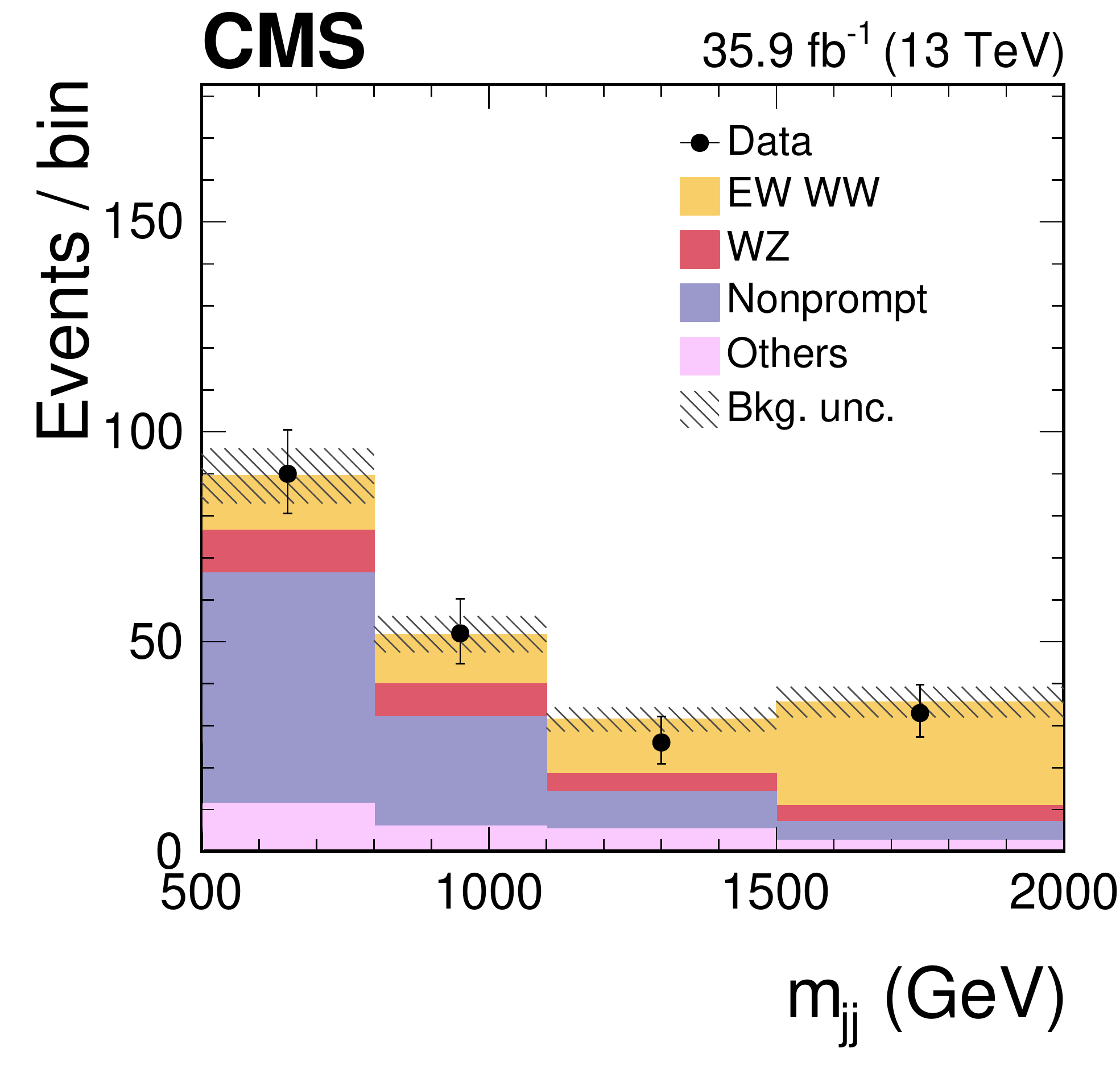}
\includegraphics[width=\cmsFigWidth]{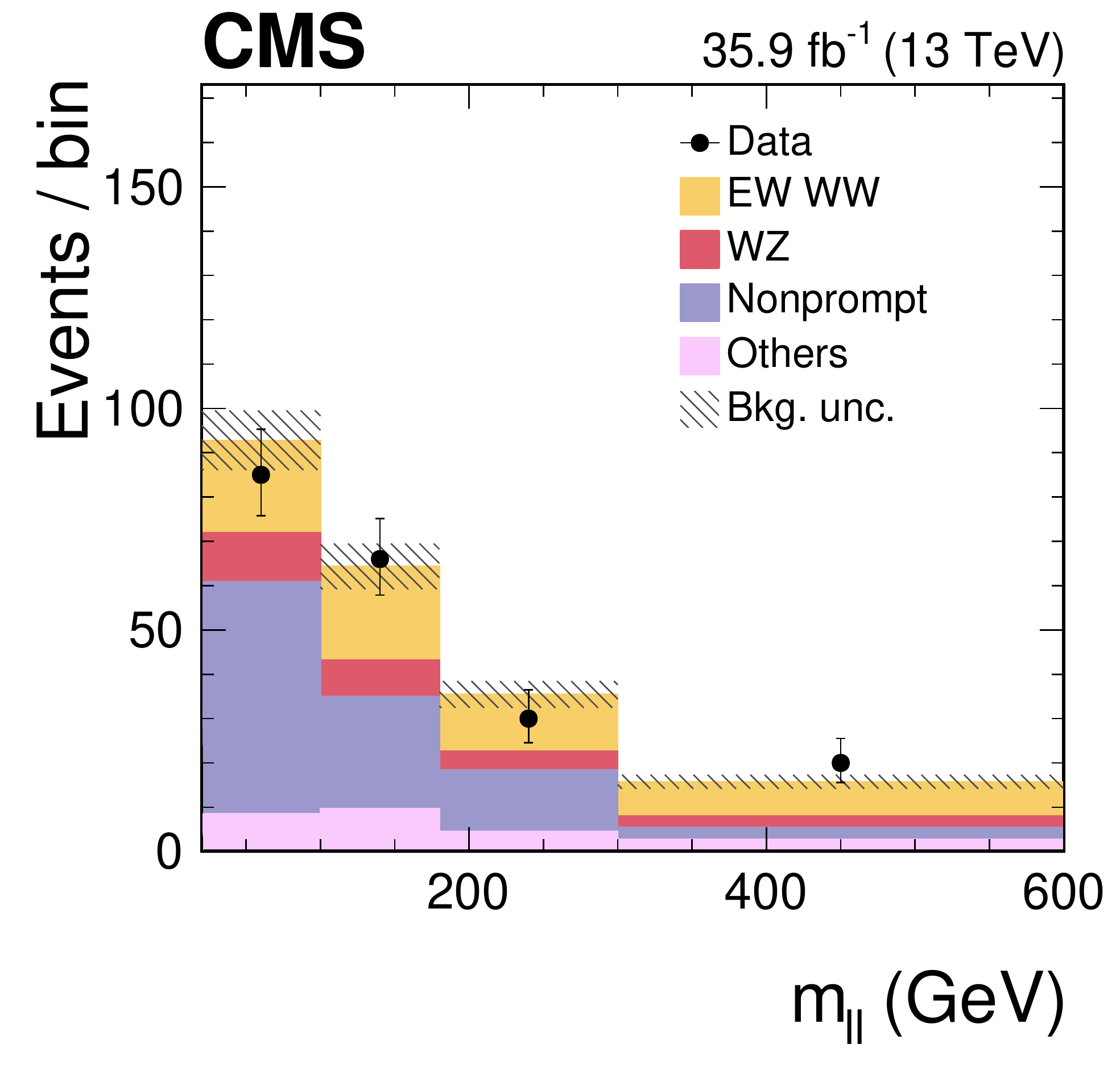}
\caption{Distributions of $m_{jj}$ (left) and $\mll$ (right) in
the signal region. The normalization of the EW $\PW^\pm\PW^\pm$ and
background distributions corresponds to the result of the fit.
The hatched bands include statistical and systematic
uncertainties from the predicted yields.
The histograms for other backgrounds include the contributions
from QCD $\PW\PW$, $\PW\gamma$, wrong-sign events, double parton
scattering, and triboson processes. The overflow is included in
the last bin.}
\label{fig:signal}
\end{figure*}

The simulated signal and background yields, as well as the observed data
yields, are shown in Table~\ref{tab:yields}.
See \suppMaterial{} for a table with more detailed results.
The two dominant sources of background events arise from nonprompt leptons and  the $\PW\PZ$ process.
The distributions of $m_{jj}$ and $\mll$ in the signal region are shown in Fig.~\ref{fig:signal}.
An excess of events with respect to the background-only hypothesis is
observed.
In order to quantify the significance of the observation of the EW
production of same-sign $\PW$ boson pairs, a statistical analysis of the
event yields is performed with a fit to the ($m_{jj}$,$\mll$) two-dimensional distributions.
The fit is performed simultaneously in the signal region and in the
$\PW\PZ$ control region, although only the $m_{jj}$
distribution is used in the latter region. The aim of using the $\PW\PZ$ control region is
to determine the number of $\PW\PZ$ background events in the signal
region as a function of $m_{jj}$.
The lepton flavor is not used to separate event
samples. The EW signal yield and the $\PW\PZ$
background normalization are free parameters of the fit. All background contributions can vary within the estimated
uncertainties. The data excess is quantified by calculating the
p-value using a profile likelihood ratio test
statistic~\cite{Junk,Read,CLs}. The observed (expected) statistical
significance of the signal is 5.5 (5.7) standard deviations.
The ratio of measured signal event yield to that expected from the SM is $0.90\pm0.22$.

\begin{table}[htb]
\centering
\topcaption{Estimated signal and background yields after the selection. The statistical and systematic
uncertainties are added in quadrature. The processes contributing to less
than 1\% of the total background are not listed, but included in the
total background yield.}
\label{tab:yields}
\begin{scotch}{lc}
Data                   &   201                  \\
\hline
Signal + total bkg.      &  205 $\pm$  13 \\
\hline
Signal                 &   66.9 $\pm$   2.4 \\
\hline
Total bkg.             &  138 $\pm$  13 \\
\hline
Nonprompt             &   88 $\pm$  13 \\
$\PW\PZ$                 &   25.1 $\pm$   1.1 \\
QCD $\PW\PW$           &   4.8 $\pm$   0.4 \\
$\PW\gamma$          &   8.3 $\pm$   1.6 \\
Triboson               &   5.8 $\pm$   0.8 \\
Wrong sign             &  5.2 $\pm$   1.1 \\
\end{scotch}
\end{table}

The cross section is extracted in a fiducial signal region,
defined using MC generator quantities
by requiring  two same-sign leptons from $\PW$ boson decays with
$\pt^{\ell}>20\GeV$ and $\abs{\eta_{\ell}}<2.5$, two jets with $\pt^{j}>30\GeV$
and $\abs{\eta^{j}}<5.0$, $m_{jj}>500\GeV$, and  $\abs{\Delta \eta_{jj}}>2.5$.
In this definition, the leptons are defined at particle level post final state
radiation and $\PW \to \PGt\PGn \to \ell\PGn\PGn\PGn$ decays are excluded.
The measured cross section is corrected for the
acceptance in this region using the \MADGRAPH{}5\_a\MCATNLO generator,
which is also used to estimate the theoretical cross section at LO.
The fiducial cross section is measured to be $\sigma_\text{fid}(\PW^{\pm}\PW^{\pm} jj) =
3.83 \pm 0.66\stat \pm 0.35\syst\unit{fb}$.
The predicted theoretical cross section at LO is $ 4.25 \pm 0.27\unit{fb}$, in agreement with the measurement. The uncertainty in the
theoretical cross section stems from scale variations and parton
distribution functions. Complete NLO QCD and EW corrections to $\PW^+\PW^+$
scattering~\cite{Biedermann:2017bss} are computed using similar selection requirements as
presented in this paper. The NLO EW corrections to the fiducial cross section are dominant and negative ($-13\%$).
The overall efficiency within the fiducial region is
$34.8 \pm 0.3\stat \pm 2.3\syst \%$.

Various extensions of the SM alter the couplings between vector bosons.
Reference~\cite{aqgc_operators} proposes nine independent charge
conjugate and parity-conserving dimension-eight
effective operators to modify the quartic couplings. In this case, the
$\mll$ distributions in both the signal and
$\PW\PZ$ regions are used to perform the statistical analysis. The EW
production is treated as a background consistent with the SM
expectation and can vary within the estimated uncertainties.
The observed and expected 95\% confidence level (CL) limits for the nine coefficients, shown
in Table~\ref{tab:VBS_aQGC}, are obtained by varying
the effective operators one by one.
The effect of possible aQGCs on the $\PW\PZ$ process in
the signal region is negligible because the background is normalized
using data.
The table also shows the most stringent 95\% CL
limits reported by the CMS Collaboration previously.

\begin{table}[h!]
\centering
\topcaption{
Observed and expected 95\% CL limits on the coefficients
for higher-order (dimension-eight) operators in the effective
field theory Lagrangian.
}
\label{tab:VBS_aQGC}
\begin{scotch}{ccc}
\ifthenelse{\boolean{cms@external}}{
& Observed  & Expected  \\
& limits  & limits \\
}
{
& Observed limits  & Expected limits  \\
}
& (\TeV$^{-4}$)   & (\TeV$^{-4}$)   \\
\hline
$\mathrm{f_{S0}} / \Lambda^4$  & $[ -7.7, 7.7]$ & $[ -7.0, 7.2]$ \\
$\mathrm{f_{S1}} / \Lambda^4$  & $[-21.6, 21.8]$ & $[-19.9, 20.2]$ \\
$\mathrm{f_{M0}} / \Lambda^4$  & $[-6.0, 5.9]$ & $[-5.6, 5.5]$ \\
$\mathrm{f_{M1}} / \Lambda^4$  & $[ -8.7, 9.1]$ & $[ -7.9, 8.5]$ \\
$\mathrm{f_{M6}} / \Lambda^4$  & $[-11.9, 11.8]$ & $[-11.1, 11.0]$ \\
$\mathrm{f_{M7}} / \Lambda^4$  & $[-13.3, 12.9]$ & $[-12.4, 11.8]$ \\
$\mathrm{f_{T0}} / \Lambda^4$  & $[-0.62, 0.65]$ & $[-0.58, 0.61]$ \\
$\mathrm{f_{T1}} / \Lambda^4$  & $[-0.28, 0.31]$ & $[-0.26, 0.29]$ \\
$\mathrm{f_{T2}} / \Lambda^4$  & $[-0.89, 1.02]$ & $[-0.80, 0.95]$ \\
\end{scotch}
\end{table}

Doubly charged Higgs bosons are predicted in models that contain a
Higgs triplet field. Some of these scenarios predict same-sign lepton events from $\PW^\pm\PW^\pm$
decays with a VBF topology. The Georgi--Machacek model of Higgs
triplets~\cite{GMH} is considered.  The couplings depend on
$m_{\PH^{\pm\pm}}$ and the parameter $\sin \theta_{\PH}$, or $s_{\PH}$,
where $s_{\PH}^2$ denotes the fraction of the $\PW$ boson mass
generated by the vacuum expectation value of the triplets.
The  expected signal event yields for VBF production of $\PH^{\pm \pm}$ decaying to
$\PW^\pm\PW^\pm$ is directly proportional to $s_{\PH}^2$.
The remaining five parameters in the
model are adjusted to achieve the given $m_{\PH^{\pm \pm}}$
hypothesis, while requiring one of the scalar singlets to have a
mass of 125\GeV.
By using the ($m_{jj},\mll$) two-dimensional distribution in the
signal region and the $m_{jj}$ distribution in the $\PW\PZ$ control
region simultaneously to discriminate between signal and background processes, 95\% CL upper
limits on $\sigma_{\mathrm{VBF}}(\PH^{\pm\pm}) \, \mathcal{
B}(\PH^{\pm\pm} \to \PW^{\pm}\PW^{\pm})$ can be derived, as shown in
Fig.~\ref{fig:ana_hpp_samesign_logx0_logy0}. The observed limit excludes $s_{\PH}$
values greater than 0.18 and 0.44 at $m(\PH^{\pm\pm}) = 200$ and 1000\GeV, respectively.
See Supplemental Material at [URL] for the expected and observed 95\% CL upper
limits on $s_{\PH}$ in the Georgi--Machacek model as a function of
doubly charged Higgs boson mass.

\begin{figure}[htb]
\centering
\includegraphics[width=\cmsFigWidth]{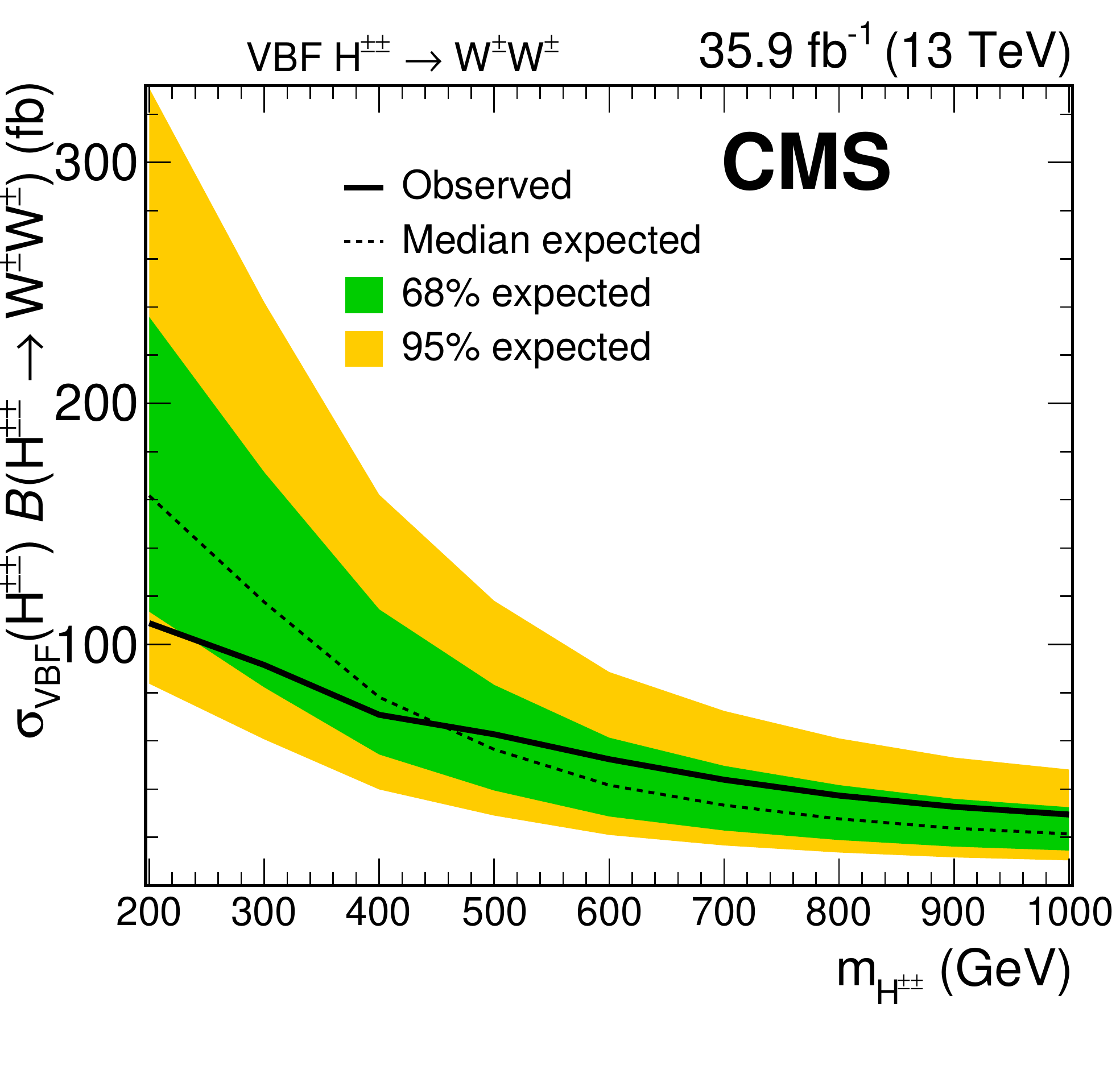}
\caption{Expected and observed 95\% CL upper
limits on the cross section times branching fraction,
$\sigma_{\mathrm{VBF}}(\PH^{\pm\pm}) \, \mathcal{B}(\PH^{\pm\pm}
\to \PW^{\pm}\PW^{\pm})$ as a function of
doubly charged Higgs boson mass.}
\label{fig:ana_hpp_samesign_logx0_logy0}
\end{figure}

In summary, we present the first observation of electroweak production
of same-sign $\PW$ boson pairs in proton-proton collisions.
The data sample corresponds to an integrated luminosity of 35.9\fbinv
collected at $\sqrt{s}=13\TeV$ with the CMS detector. Events are selected by requiring
exactly two leptons of the same charge, moderate $\ptmiss$, and two jets with large rapidity separation and large dijet mass. The two dominant sources of background events after the event
selection requirements have been applied are nonprompt leptons and  the $\PW\PZ\to 3\ell\PGn$ process.
The observed significance is 5.5 standard deviations, where a
significance of 5.7 standard deviations is expected based on the
SM. The ratio of measured event yields to that expected from the standard model at leading-order
is $0.90\pm 0.22$. A cross section measurement in a fiducial region is
reported consistent with SM predictions.
Bounds on the structure of quartic vector boson interactions are improved by a factor of up to six compared to previous results. Upper limits are given on the production cross section times branching fraction of doubly charged Higgs bosons.

\begin{acknowledgments}
We congratulate our colleagues in the CERN accelerator departments for the excellent performance of the LHC and thank the technical and administrative staffs at CERN and at other CMS institutes for their contributions to the success of the CMS effort. In addition, we gratefully acknowledge the computing centres and personnel of the Worldwide LHC Computing Grid for delivering so effectively the computing infrastructure essential to our analyses. Finally, we acknowledge the enduring support for the construction and operation of the LHC and the CMS detector provided by the following funding agencies: BMWFW and FWF (Austria); FNRS and FWO (Belgium); CNPq, CAPES, FAPERJ, and FAPESP (Brazil); MES (Bulgaria); CERN; CAS, MoST, and NSFC (China); COLCIENCIAS (Colombia); MSES and CSF (Croatia); RPF (Cyprus); SENESCYT (Ecuador); MoER, ERC IUT, and ERDF (Estonia); Academy of Finland, MEC, and HIP (Finland); CEA and CNRS/IN2P3 (France); BMBF, DFG, and HGF (Germany); GSRT (Greece); OTKA and NIH (Hungary); DAE and DST (India); IPM (Iran); SFI (Ireland); INFN (Italy); MSIP and NRF (Republic of Korea); LAS (Lithuania); MOE and UM (Malaysia); BUAP, CINVESTAV, CONACYT, LNS, SEP, and UASLP-FAI (Mexico); MBIE (New Zealand); PAEC (Pakistan); MSHE and NSC (Poland); FCT (Portugal); JINR (Dubna); MON, RosAtom, RAS, RFBR and RAEP (Russia); MESTD (Serbia); SEIDI, CPAN, PCTI and FEDER (Spain); Swiss Funding Agencies (Switzerland); MST (Taipei); ThEPCenter, IPST, STAR, and NSTDA (Thailand); TUBITAK and TAEK (Turkey); NASU and SFFR (Ukraine); STFC (United Kingdom); DOE and NSF (USA).
\end{acknowledgments}

\bibliography{auto_generated}
\ifthenelse{\boolean{cms@external}}{}{
\clearpage
\appendix
\section{Supplemental material\label{app:suppMat}}
\begin{table*}[htbp]
\begin{center}
\topcaption{Estimated signal and background yields after the selection. The statistical
uncertainties are reported for all six channels, while
the sums are reported with the statistical and systematic
uncertainties added in quadrature. The processes contributing to less
than 1\% of the total background are not listed, but included in the
total background yield.}
\label{tab:yields}
\ifthenelse{\boolean{cms@external}}{}{\resizebox{\textwidth}{!}}
{
\begin{scotch}{lccccccc}
& $\Pep\Pep$            &	$\Pep\Pgmp$        &  $\Pgmp\Pgmp$      &	$\Pem\Pem$            &	$\Pem\Pgmm$        &  $\Pgmm\Pgmm$         &  Total		  \\
\hline
Data                   &   14                &             63        &  40                &      10                 & 48                  &             26        &  201                  \\
\hline
Signal + total bkg.      &  19.0 $\pm$   1.9\x &    67.6 $\pm$   3.8\x & 44.1 $\pm$   3.4\x &    11.8 $\pm$  1.8\x  &  38.9 $\pm$	3.3\x &    23.9 $\pm$   2.8\x &	205 $\pm$  13\x \\
\hline
Signal                 &   6.2 $\pm$   0.2 &    24.7 $\pm$   0.4\x & 18.3 $\pm$   0.4\x &     2.5 $\pm$   0.1 &   8.7 $\pm$	0.2 &     6.5 $\pm$   0.2 &	 66.9 $\pm$   2.4\x \\
\hline
Total bkg.             &  12.8 $\pm$   1.9\x &    42.9 $\pm$   3.8\x & 25.7 $\pm$   3.4\x &     9.4 $\pm$   1.8 &  30.2 $\pm$	3.3\x &    17.4 $\pm$   2.8\x &	138 $\pm$  13\x \\
\hline
Nonprompt             &   5.6 $\pm$   1.7  &   24.9 $\pm$	3.6\x & 18.4 $\pm$   3.3\x &     5.0 $\pm$   1.6 &   19.9 $\pm$  3.2\x &	 14.2 $\pm$   2.8\x &	 88 $\pm$  13 \\
$\PW\PZ$                 &   3.0 $\pm$   0.2 &     8.5 $\pm$   0.3 &  4.4 $\pm$   0.2 &     1.9 $\pm$   0.2 &    5.2 $\pm$  0.3 &	  2.2 $\pm$   0.1 &	 25.1 $\pm$   1.1\x \\
QCD $\PW\PW$           &   0.6 $\pm$   0.1 &     1.7 $\pm$   0.1   &  1.3 $\pm$   0.1 &     0.2 $\pm$   0.1 &    0.6 $\pm$  0.1 &	  0.4 $\pm$   0.1 &	  4.8 $\pm$   0.4 \\
$\PW\gamma$          &   1.4 $\pm$   0.5 &     3.6 $\pm$   0.9   &  0.2 $\pm$   0.2 &     0.8 $\pm$   0.4 &    2.3 $\pm$  0.7 &      \NA        &	  8.3 $\pm$   1.6 \\
Triboson               &   0.8 $\pm$   0.2 &     2.2 $\pm$   0.4 &  1.2 $\pm$   0.3 &	0.3 $\pm$   0.1 &    0.9 $\pm$  0.3 &     0.5 $\pm$   0.2 &	  5.8 $\pm$   0.8 \\
Wrong sign             & 1.5 $\pm$   0.6 &     1.4 $\pm$   0.4  & \NA       &	1.1 $\pm$   0.5 &    1.2 $\pm$  0.4     &   \NA           &	  5.2 $\pm$   1.1 \\
\end{scotch}
}
\end{center}
\end{table*}

\begin{figure}[htbp]
\begin{center}
\includegraphics[width=\cmsFigWidth]{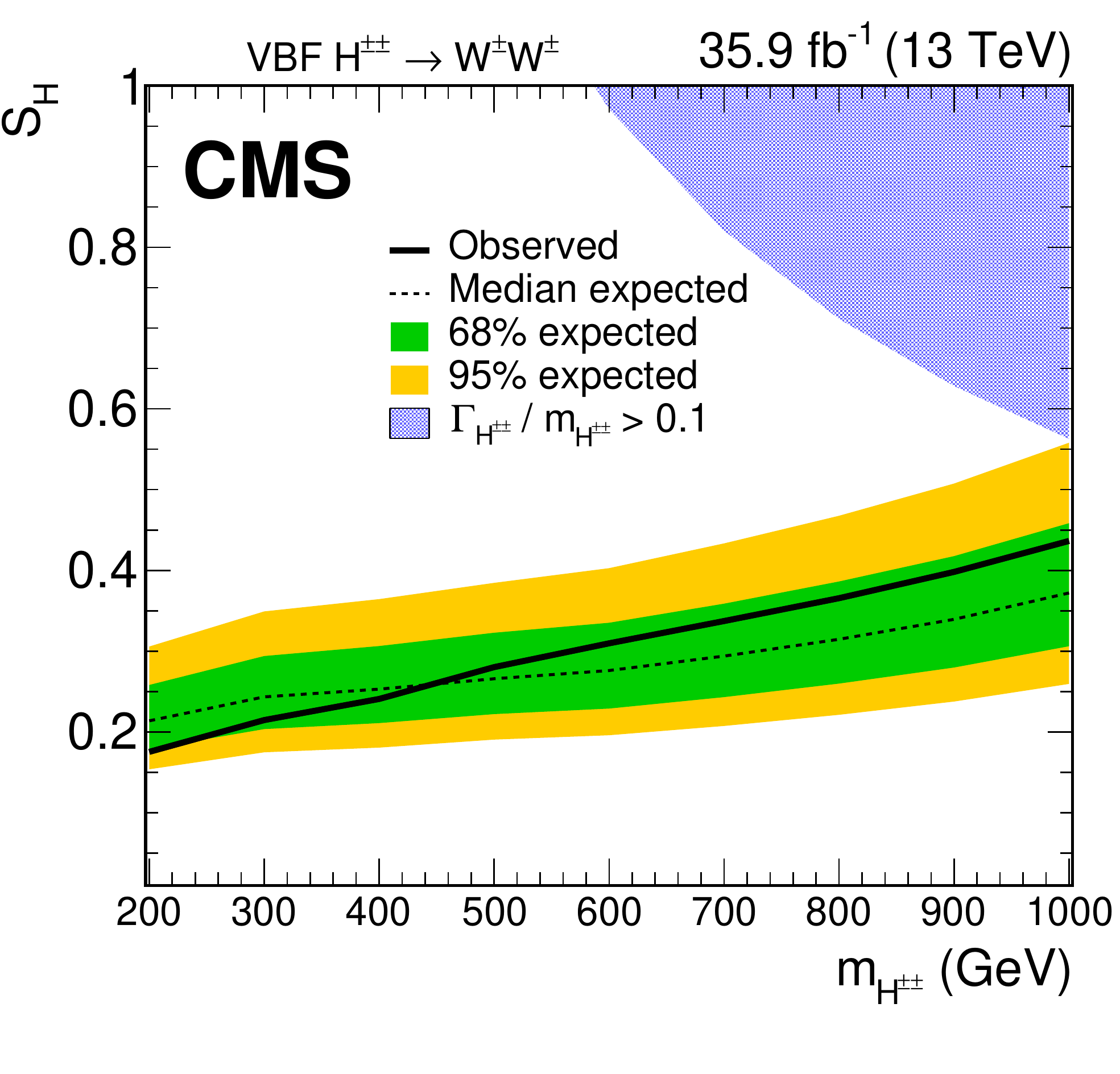}
\caption{Expected and observed 95\% CL upper
limits on $s_{\PH}$ in the Georgi--Machacek model as a function of
doubly charged Higgs boson mass.
The blue area in the upper-right corner covers the region where the model is not applicable~\cite{Zaro:2002500}.}
\label{fig:ana_hpp_samesign_logx0_logy0}
\end{center}
\end{figure}

}

\cleardoublepage \section{The CMS Collaboration \label{app:collab}}\begin{sloppypar}\hyphenpenalty=5000\widowpenalty=500\clubpenalty=5000\textbf{Yerevan Physics Institute,  Yerevan,  Armenia}\\*[0pt]
A.M.~Sirunyan, A.~Tumasyan
\vskip\cmsinstskip
\textbf{Institut f\"{u}r Hochenergiephysik,  Wien,  Austria}\\*[0pt]
W.~Adam, F.~Ambrogi, E.~Asilar, T.~Bergauer, J.~Brandstetter, E.~Brondolin, M.~Dragicevic, J.~Er\"{o}, M.~Flechl, M.~Friedl, R.~Fr\"{u}hwirth\cmsAuthorMark{1}, V.M.~Ghete, J.~Grossmann, J.~Hrubec, M.~Jeitler\cmsAuthorMark{1}, A.~K\"{o}nig, N.~Krammer, I.~Kr\"{a}tschmer, D.~Liko, T.~Madlener, I.~Mikulec, E.~Pree, N.~Rad, H.~Rohringer, J.~Schieck\cmsAuthorMark{1}, R.~Sch\"{o}fbeck, M.~Spanring, D.~Spitzbart, W.~Waltenberger, J.~Wittmann, C.-E.~Wulz\cmsAuthorMark{1}, M.~Zarucki
% ...
\vskip\cmsinstskip
% ...
1:~~Also at Vienna University of Technology, Vienna, Austria\\
% ...

\end{sloppypar}
\end{document}